# A New Approach for Macroscopic Analysis to Improve the Technical and Economic Impacts of Urban Interchanges on Traffic Networks

Seyed Hassan Hosseini[1] [0000-0003-3239-3163] Ahmad Mehrabian[2] [1111-2222-3333-4444], Zhila Dehdari Ebrahimi[3][0000-0001-7256-0881], Mohsen Momenitabar[4][0000-0003-2568-1781] and Mohammad Arani[5][0000-0002-1712-067X]

[1] Sapienza University of Roma, Department of Civil and Industrial Engineering, Roma, Italy
`hosseini.1851467@studenti.uniroma1.it`
[2] Islamic Azad University, Department of Industrial Engineering, Aliabad Katoul Branch, Iran
`ahmad.mehrabian@outlook.com`
[3] College of Business, North Dakota State University, Fargo, ND 58102, USA
`Zhila.dehdari@ndsu.edu`
[4] College of Business, North Dakota State University, Fargo, ND 58102, USA
`mohsen.momenitabar@ndsu.edu`
[5] University of Arkansas at Little Rock, 2801 S. University Ave., Little Rock, AR 72204, USA
`mxarani@ualr.edu`

**Abstract.** Pursuing three important elements including economic, safety, and traffic are the overall objective of decision evaluation across all transport projects. In this study, we investigate the feasibility of the development of city interchanges and road connections for network users. To achieve this goal, a series of minor goals are required to be met in advance including determining benefits, costs of implementing new highway interchanges, quantifying the effective parameters, the increase in fuel consumption, the reduction in travel time, and finally influence on travel speed. In this study, geometric advancement of Hakim highway, and Yadegar-e-Emam Highway were investigated in the Macro view from the cloverleaf intersection with a low capacity to a three-level directional intersection of the enhanced cloverleaf. For this purpose, the simulation was done by EMME software of INRO Company. The results of the method were evaluated by the objective of net present value (NPV), and the benefit and cost of each one was stated precisely in different years. At the end, some suggestion has been provided.

## 1 Introduction

Annually, billions of dollars are spent on traffic performance improvement, smoothing traffic flow, and reducing travel time in metropolises (Bigdeli Rad and Bigdeli Rad 2018). Construction of a grade-separated interchange is a high-cost project, which takes place for physical separation of traffic routes, continuous traffic flow, decreasing delay time, and increasing capacity of intersections. However, evidence has shown that the construction of a grade-separated interchange can increase travel time and delay time in the adjacent intersection (Khan and Anderson 2016). So, choosing the best type of interchange under different traffic conditions can greatly mitigate this adverse effect (Arani et al. 2020a; Arani et al. 2020d; Arani et al. 2020b; Arani et al. 2020c; Arani et al. 2019; Chan et al. 2020; Momeni Tabar et al. 2014; MomeniTabar et al. 2020; Sutherland et al. 2018; Zhila Dehdari Ebrahimi, Raj Bridgelall 2020).

One of the most important processes of planning is evaluating several different options to achieve a common goal to ensure that the best option is chosen, and other options are excluded. Increasing the financial value of the project leads to the increasing importance of economic evaluation (Mansourkhaki and Ghanad 2014). The purpose of constructing interchanges is to creating levels through which vehicles can move in different directions without collisions with other vehicles. Traffic capacity per line decreases in at-grade interchanges (Lucietti et al. 2016). Constructing interchanges with different levels, are the best, safest, and fastest and at the same time the most expensive solution to passing cross directions. One of the ways to increase the capacity of interchanges is by promoting the geometric characteristics of interchanges (Afandi Zade et al. 2019; Chowdhury 2016). An interchange can be an adaptable solution to improve many intersection conditions either by reducing existing traffic bottlenecks or by reducing crash frequency (Mollu et al. 2017). However, the high cost of constructing an interchange limits its use to those cases where an additional expenditure can be justified. Enumeration of the specific conditions or warrants justifying an interchange at a given intersection is difficult, and in some instances, cannot be conclusively stated. Because of the wide variety of site conditions, traffic volumes, highway types, and interchange layouts, the warrants that justify an interchange may differ at each location. The following six conditions, or warrants, should be considered when determining if an interchange is justifiable at a particular site (of State Highway and Officials 2011): Design designation, Reduction of bottlenecks or congested spot, Reduction of collisions frequency and severity, Site topography, Road-user benefits, and Traffic volume warrant.

The first cloverleaf was opened in Woodbridge Township, New Jersey in 1929, at the junction of Routes 4 and 25. This is now the intersection of US 1/9 and NJ 35, and since about 2004, the interchange is no longer a cloverleaf. The classic cloverleaf allows "non-stop" full access between two busy roads. Traffic merges and weaves but does not cross at-grade; unless the interchange is heavily congested, no stop is required. The colloquial "cloverleaf" is the same as the technical "full cloverleaf", as one can omit ramps to get a partial one. The cloverleaf, which is studies here, is the simplest way to connecting two freeways. Bridges are only required to separate two roadways. If the land is expensive, which becomes a choice between tight turning radii (and lower design speed). Note that most loop ramps are banked to counteract centrifugal forces (of State Highway and Officials 2011). Typically, a cloverleaf is used where a freeway intersects a busy surface street, though many older freeway interchanges are also con-sidered cloverleaves (Chan et al. 2020). In several places, cloverleaves have been replaced with either signalized interchanges or higher-capacity directional interchanges with flyovers.

## 2 Method

Implementation and exploitation of each transportation plan in road networks can have a positive effect on transport and traffic flow of a city, region, or area, resulting in improved traffic conditions. But what is of considerable importance is comparing the benefits and costs of implementing the plan to assess its economic value. In this study, a common method is presented for the economic evaluation of the implementation of a traffic plan as a grade-separated interchange. According to **Fig. 1**, in this study, there is a new way of looking at the highway interchanges, which is a macroscopic view. Thus, it is expressed as an element of the urban transportation network (Zhila Dehdari Ebrahimi 2017) with a very broad sphere of influence and effects of trans-regional and macroscopic view. Therefore, two logical scenarios were defined and after the simulation in macroscopic software EMME, network information was obtained. To compare the parameters of travel time, safety, environment, and fuel, these parameters are converted to economic values (dollars) to determine to what extent the plan is justifiable.

In this study, using the statistical models based on Tehran Transportation and Traffic Studies and employing EMME, the macroeconomic software, a wide variety of effects regarding the changes in the highway network was investigated. In this case, for 15 years, these effects on the highway network, due to the increase in the population and ownership of the car, and public transportation were reviewed.

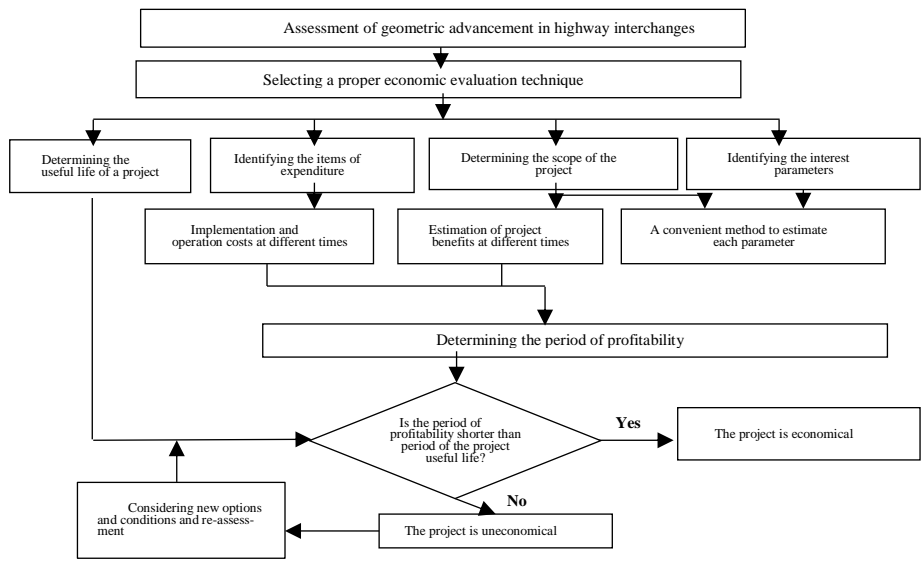

**Fig. 1.** The methodology of the study

Previous studies have shown that these long-term changes could be beneficial or possibly harmful to the community. On the other hand, the point that is often neglected in this assessment (despite the great importance) of highway exchange projects is the costs incurred by users of the urban transportation network during the execution of such projects. These costs include costs of inaccessibility, increasing the length of the route, changing certain routes that were used in the past, land use, etc. Typically, traffic peak-hours at 2 stages in the morning and night are the basis for designing the traffic load. But the very important point is that in analyzing and evaluating projects, it should not be evaluated economically merely according to the peak hours of the city. Due to the low traffic volume of vehicles, the time spent, and the increase in fuel consumption, there is no increase in environmental pollution. So, as the peak-based design is an over-estimated design, it can be useful, but the economic assessment based on peak hours of the city has a high error rate and is somewhat irrefutable. But in the current research methodology, the evaluation is based on 24 hours a day, which indicates high accuracy. The contribution of this paper can be seen in **Table 1**.

**Table 1.** Summary of contribution (Baldauf et al. 2013)

| Benefits | Aim | Method | Software |
|---|---|---|---|
| Travel time reduction | Estimating travel time in both the existing and the proposed scenarios | Macro simulation in the transportation system | EMME |
| Fuel consumption reduction | Estimating fuel consumption in both the existing and the proposed scenarios | Hicks and Clarkson Model | EMME |
| Air pollution reduction | Estimating various types of air pollutants emissions in both the existing and the proposed scenarios | Calibrated models in the laboratories of fuel consumption optimization organization | EMME |
| Collision reduction | Estimating the number of collisions | A linear relationship between speed and collisions | - |

In this study, the method of "net present value" is recommended to assess the economic value of intersections. Net Present Value (NPV) is a formula used to determine the present value of an investment by the sum of all cash flows received from the project over time (Gonzalez-Ruiz et al. 2017). The formula for the sum of all cash flows can be rewritten as:

$$NPV = -C_0 + \sum_{i=1}^{T} \frac{C_i}{(1+r)^i} \tag{1}$$

When a company or investor takes on a project or investment, it is important to calculate an estimate of how profitable the project or investment would be (Gonzalez-Ruiz et al. 2017). NPV is equal to the equivalent value of the profits from the investment regarding the start time of the project, where the value of operational costs, maintenance cost, and cost of final equipment exploitation is considered by the monetary value of currency over time (Pfeiffer 2004). Based on economic principles, exploitation of an at-grade separated intersection would be economically justified when the desired profitability is reached before the end of its lifetime (Madanu et al. 2010). Costs and benefits of the project in the years after the start of exploitation will be the basis for determining the NPV index, in which it will be determined with software EMME. To evaluate the effectiveness of the proposed method, two different designs of at-grade separated intersections were considered at the current intersection. The proposed plan of alternative intersection is outlined in **Fig. 3**, which is a three-level directional intersection, where all the turnings are grade-separated and independent, which will be done with high speed. The movements are the two left-turn movements and the major routes of Mohammad Ali Jinnah in the underpass of Hakim Highway Bridge, the turn-right movement on the ground floor, and the two left-turn movements on the upper floor.

According to the proposed plan, the intersection is combined with six bridges where two of them are the main link between the eastern and western parts of the city and the other four bridges are for the ramp. Including improved areas in the main highways, this plan has about 8 KM circulation paths and ramp, $25000 m^2$ bridges and more than $1000 m^2$ walls.

The current plan of Hakim and Yadegar-e Emam highway is a cloverleaf intersection (**Fig. 3**), but the northwest ramp has occupied a very large area, increasing the length of the route and reduces driver sight distance in parts of the route. Further, at this intersection due to the low width at the main bridge, ramps, and loops, current demand cannot be satisfied. There will be heavy congestion during peak hours. In this plan, the current cloverleaf intersection will expand, and the width of ramps, loops, and bridges will increase. The geometric design of the northwest loop will become a standard loop with appropriate sight distance and the enhanced cloverleaf intersection will develop with more capacity. This plan has been shown in **Fig. 2**.

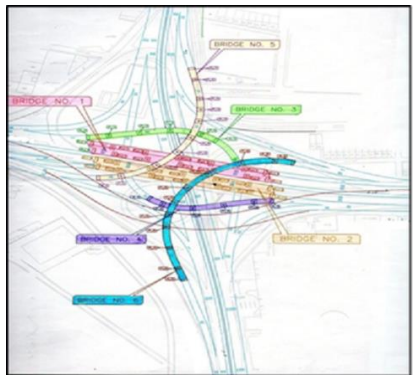
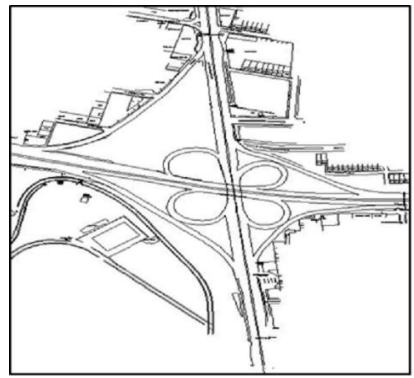

**Fig. 2**. Proposed plan for directional interchange versus status quo

**Fig. 3.** The proposed plan for cloverleaf interchange versus status quo

As the simulation process in the current situation, in the construction phase and years after employment is the most important step in the estimation of costs imposed on citizens during the construction as well as the benefits arising from the utilization of the intersection. This section deals with the simulation process of transportation systems and traffic conditions in Tehran, implemented in EMME software.

**Fig. 4** demonstrates the general trend of demand estimated for internal trips of residents in Tehran. The demand estimation process of other trips of residents and non-residents travel is shown in **Fig. 5**. According to **Fig. 4**, at first, using calibrated models of generation and attraction, the daily trips of 560 traffic zones in Tehran are predicted. The prediction is according to non-home-based trips and home base trips (7).

**Table 2.** Trip generation and attraction models of Tehran for each trip purpose

| Trip purpose | Trip attraction models |
|---|---|
| Work | $TA_i^w = 1/620 EMPE_i + 2/420 SHOP_i + 62694 DB_i$ |
| Education | $TA_i^s = 3/833 VP_i * ST_i + 0/500 STU_i + 26789 DT_i + 9299 D_{128i}$ |
| Buy | $TA_i^{sh} = 15/760 VP_i * SHOP_i + 0/195 EMPE_i + 0/825 HOSB_i + 15456 DB_i + 3469 DQ_i + 7474 DF_i + 4607 D_{444i} - 0/889 SHOP_i * DRA_i$ |
| Entertainment and other | $TA_i^{sr} = 122/140 PARK_i + 0/040 P_i + 7/364 VP_i * SHOP_i + 0/304 EMPE_i + 4098 DR_i + 1937 DF_i + 1532 DQ_i - 0/279 SHOP_i * DRA_i - 0/208 EMPE_i * DRA_i$ |
| Non-home base | $TA_i^{nhb} = 0/458 EMPE_i + 11/526 VP_i * SHOP_i + 11706 DB_i + 1173 DQ_i$ |

$P_i$ = Traffic zone population
$VP_i$ = Per capita ownership of private car in traffic zone $i$
$ER_i$ = Resident employment in traffic zone $i$
$STR_i$ = The number of resident school students in traffic zone $i$
$STUR_i$ = The number of resident university students in traffic zone $i$
$EMPE_i$ = The number of staff in place of employment in traffic zone $i$
$SHOP_i$ = The number of shops in traffic zone $i$
$ST_i$ = The number of students at school in traffic zone $i$
$STU_i$ = The number of students at university in traffic zone $i$
$HOSPB_i$ = The number of hospital beds in traffic zone $i$

$PARK_i$ = The number of parks in traffic zone $i$
$DRA_i$ = Traffic zones covariate in traffic plan zone
$DB_i$ = Tehran Bazaar covariate (Traffic zone 1)
$DT_i$ = Tehran university covariate (Traffic zone 150)
$DQ_i$ = Main squares covariate (Imam Hossein, Enghelab, valiasr, Khorasan, Tajrish, and Imam Khomeini) in traffic zones 16, 174, 151, 121, 401, 537.
$DF_i$ = Covariate of Ghezel Ghale square, second Square of Sadeghieh mall, second mall of Nazi Abad in traffic zones 197, 232, 139.
$DR_i$ = Covariate of Mellat park, laleh park, Shahbdalzym and Behesht Zahra in traffic zones 452, 274, 148, and 444.
$D_{128}$ = Covariate of Amirkabir university, Art university and Alborz High School (Traffic zone 128)
$D_{444}$ = Covariate of Ray city (Traffic zone 444)

Due to trip purpose, some important vehicles are imported in the modeling. Note that the important vehicles are those vehicles with a major contribution to travel (9). **Table 2** shows the calibration result of the trip generation and attraction model, and **Table 3**, **Table 4**, **Table 5**, **Table 6**, **Table 7**, and **Table 8** provide the result of vehicle choice model calibration in Tehran. Vehicle choice models are Logit Models in this study.

**Table 3.** Trip Purpose

| Trip purpose | Trip generation models |
|---|---|
| Work | $T_i^w = 0/569 VP_i * ER_i + 1/107 ER_i$ |
| Education | $T_i^s = 3/070 VP_i * STR_i + 0/903 STUR_i + 0/020 DIST_i * P_i$ |
| Buy | $T_i^{sh} = 0/061 P_i + 0/414 VP_i * P_i$ |
| Entertainment | $T_i^{sr} = 0/073 P_i + 0/373 VP_i * P_i$ |
| Non-home base | $T_i^{nhb} = 0/490 EMPE_i + 10/213 VP_i * SHOP_i + 14485 DB_i + 951 DQ_i$ |

**Table 4.** Vehicle choice models of Tehran for each trip purpose

| | | **Work trip** |
|---|---|---|
| First level | Car | $U_{CAR}^{ij} = 0.697568 - 0.034575 \times TIMCAR^{ij} + 9.008179 \times owncar^i - 0.588848 \times DESFLAG^j$ |
| | Motorcycle | $U_{MOT}^{ij} = -0.480759 - 0.047222 \times TIMMOT^{ij} + 18.345253 \times OWNMOT^i$ |
| | Public transportation | $U_{BUS,TAX}^{ij} = \theta \ln[\exp(U_{BUS}^{ij}) + EXP(U_{TAX}^{ij})]$ |
| Second level | Bus | $U_{BUS}^{ij} = 0.330393 - 0.020389 \times TIMIN^{BUS} - 0.026496 \times TIMBOT^{ij}$ |
| | Taxi | $U_{TAXI}^{ij} = -0.048415 \times TIMTAX^{ij} + 3.100169 \times OWNCAR^i$ |

**Table 5.** Educational Trip

| **Educational Trip** | |
|---|---|
| Bus | $U_{BUS}^{ij} = 0.8811690 - 0.012004 \times (TIMBIN^{ij} + TIMBOT^{ij})$ |
| Taxi | $U_{TAX}^{ij} = -0.0365572 - 0.030786 \times TIMTAX^{ij} + 8.253327 \times OWNCAR^i$ |
| Car | $U_{CAR}^{ij} = -1.1044833 - 0.041592 \times TIMCAR^{ij} + 11.324764 \times owncar^i - 0.582493 \times DESFLAG^i$ |
| Minibus | $U_{MIB}^{ij} = -1.104768 \times DIST^{ij} + 6.648515 \times owncar^i$ |

**Table 6.** Shopping Trip

| **Shopping Trip** | |
|---|---|
| Bus | $U_{BUS}^{ij} = 2.794484 - 0.013595 \times TIMBIN^{ij} - 0.015329 \times TIMBOT^{ij}$ |
| Taxi | $U_{TAX}^{ij} = 1.967395 - 0.037180 \times TIMTAX^{ij} + 6.596312 \times OWNCAR^i$ |
| Car | $U_{CAR}^{ij} = -0.015029 \times TIMCAR^{ij} + 12.443686 \times OWNCAR^i - 0.689367 \times DESFLAG^j$ |

**Table 7.** Recreational Trip

| | Recreational Trip |
|---|---|
| Bus | $U_{BUS}^{ij} = 2.725886 - 0.009414 \times (TIMBIN^{ij} + TIMBOT^{ij})$ |
| Taxi | $U_{TAX}^{ij} = 2.393202 - 0.033543 \times TIMTAX^{ij} + 5.379732 \times OWNCAR^{i}$ |
| Car | $U_{CAR}^{ij} = -0.015111 \times TIMCAR^{ij} + 13.957626 \times OWNCAR^{i} - 0.374195 \times DESFLAG^{j}$ |

**Table 8.** Non-home base Trip

| | Non-home base Trip |
|---|---|
| Bus | $U_{BUS}^{ij} = 0.039002 - 0.008689 \times TIMBIN^{ij} - 0.041852 \times TIMBOT^{ij}$ |
| Taxi | $U_{TAX}^{ij} = 0.334293 - 0.020176 \times TIMTAX^{ij}$ |
| Car | $U_{CAR}^{ij} = -0.012662 \times TIMCAR^{ij} - 0.705396 \times DESFLAG^{j}$ |

The results of the deviation of demand models from various vehicles (except bus) for public transportation due to new technology are provided in **Table 9**. Note that these models are the same for all types of intercity trips of residents (all trip purposes).

**Table 9.** Results of calibration of demand deviation models

| Vehicle type | $b^m$ | $a^m$ |
|---|---|---|
| Car | 19/198 | 1/559 |
| Taxi | 2/668 | 4/794 |
| Minibus | 19/198 | 1/652 |
| Bicycle | 19/198 | 1/652 |

What is important in urban transportation planning is travel demand based on vehicles. Therefore, the demand matrix $TP_{ij}^{mt}$ must be converted to travel demand matrix based on vehicle, for all vehicles except bus. As shown in **Table 9**, this is done using information about the average number of passengers. By adding freighter trips to the mentioned matrix, the final travel demand based on the vehicle in any period $t$ can be achieved (similarly, except bus). According to the considered Passenger Car Equivalent for each vehicle, these matrixes are converted as demand matrix in terms of Passenger Car Equivalent. Passenger Car Equivalent for each vehicle is provided in Table (10).

**Table 10.** Passenger Car Equivalent

| Vehicle Type | Car | Pickups | Taxi | Minibus | Bus | Motorcycle | Lorry |
|---|---|---|---|---|---|---|---|
| PCE | 1 | 1 | 2 | 2.5 | 2.5 | 0.5 | 2.5 |

An assignment is the last step of transportation planning (UTPS) which consists of two parts. The first part is known as the car assignment. It is related to vehicles

9which have no fixed and certain route. Given the network that users can use, they try to minimize their travel time. The second part is the public transportation assignment, which is related to vehicles with a fixed and predetermined route, such as bus and subway. Users, according to a certain plan, prefer to minimize the expected travel time (Bigdeli Rad and Bigdeli Rad 2018). Based on the evidence in the Technical Organization of Tehran Municipality, the approximate costs of the items listed in each of the two proposals of the standard cloverleaf and directional interchanges are provided in **Table 11**.

**Table 11.** Direct costs of construction and operation

| Cost type | Monetary equivalent (million $) | |
|---|---|---|
| | Standard cloverleaf interchange | Directional interchange |
| Construction | 31.4 | 40.3 |
| Acquisition and release | - | 16.4 |
| Annual Maintenance | 0.55 | 0.55 |

## 3 Result

Estimation of the costs imposed on citizens during the operation and construction of interchanges will be computed according to the methods proposed in the previous chapter. Simulation of the street network of Tehran in both the status quo and network status during the operation and construction is the basis of estimating these costs. For this purpose, the transport and traffic model in Tehran will be simulated in software EMME. The changes in network performance indicators because of constraints caused by interchanges construction are demonstrated in **Table 12**.

**Table 12.** The changes in network performance

| | Index | Distance Increasing (%) | Volume reduction (%) | Pollutants emission increasing (%) | Fuel consumption increasing (Litter) | Waste-time increasing (Hour) |
|---|---|---|---|---|---|---|
| Rate of Change | Directional | 4.32 | 0.04 | 4.33 | 124325265 | 28620924 |
| | Cloverleaf | 35.8 | 0.028 | 3.51 | 98710254 | 22912723 |

Based on the method proposed in this study, the average value of travel time per person is required. For this purpose, the results related to the Tehran transportation master plan will be used, which is calculated by dividing GDP by the number of annual production hours of employed people. This index was calculated to be 2.004 dollars per hour in 2013. With increasing 28,620,924 hours in network travel time, due to construction of the directional interchange, 57.6 million dollars will be imposed on citizens. Moreover, the costs resulting from cloverleaf construction operation is estimated at 32.9 million dollars.

According to the Statistics of fuel consumption optimization organizations, fuel consumption in Tehran is 14% gasoline and 88% petrol. Due to the low percentage of gasoline and the low price, petrol is considered for all fuel consumption. The price of 1 litter petrol was equal to 0.7 dollars. Given 124,325,265 litters increase in fuel consumption caused by limitations resulting from the construction of the directional interchange, costs imposed on society are estimated at 86.6 million dollars. Also, such costs are 49.5 million dollars for constructing cloverleaf interchange.

The total direct costs resulting from accidents in Tehran are around 110 million dollars (based on the statistics presented by the Division of Insurance). Regarding value, about 4% are fatal accidents, 24% are injury accidents, and 72% are property damage accidents. With a growth of 4.32% in the traveled distances and increasing accidents, their costs are estimated to be 4.7 million dollars, which will further increase the finished cost of the project. On the other hand, the changes in traffic conditions through the black spots in the area of interest are very influential in the damage caused by accidents due to the geometric promotion of the interchange.

According to information from Tehran Traffic Police and Tehran Central Insurance Office, 3 important hot spots were determined in the plan area as demonstrated in **Fig. 7**. There is a direct relationship between the number of vehicles passing through the hot spots and the number of accidents and consequently the cost of accidents. A major change in the network leads to an increase or decrease in the capacity through the link. With estimating the traffic volume passing through these points before and after the promotion of the intended interchange, one can determine what percentage was the former traffic volume crossing the black spots and what percentage of the alternative will pass through routes or Bypass.

In 2006, the World Bank published a report stating that pollutants cost in Tehran traffic network is around 700 million dollars per year. These costs include deaths from air pollution, treatment including hospital stays, outpatient treatments, and the number of working days lost.

**Table 13.** The changes in network performance

| Iran | $NO_x$ | $SO_2$ | CO | $NMVOC_s$ |
|---|---|---|---|---|
| Value ($/tonne) | 600 | 1.825 | 188 | 0.5 |

Accordingly, concerning the 4.33% increase in pollutant emission of HC, CO, and NOx based on **Table 13**, the additional costs imposed on the society caused by

4construction operation during construction of the directional interchange are estimated at 36.6 million dollars. Under similar circumstances, 21.5 million dollars is estimated for cloverleaf interchange.

**Table 14.** The changes in network performance (Exhaust emissions)

| Pollutant | During construction (Kg) | Status quo (Kg) |
|---|---|---|
| CO | 898952107.3 | 862434217.6 |
| HC | 113958772.6 | 108149728.5 |
| $NO_x$ | 20669540.7 | 19821223.1 |

Thus, according to the description mentioned above, the total cost of the proposal of directional interchange and standard cloverleaf interchange at the beginning of the exploitation 2013 is presented to **Table 15.** Cost of the proposed plan of interchanges.

**Table 15.** Cost of the proposed plan of interchanges

| Cost type | Cost components | Cost (billion IRR) | |
|---|---|---|---|
| | | Directional interchange | Cloverleaf interchange |
| Direct | Construction | 403 | 314 |
| | Acquisition and release | 164 | - |
| As a result of limitations in the network caused by the construction process | Travel time increasing | 576 | 329 |
| | Fuel consumption increasing | 866 | 495 |
| | Pollutants emission increasing | 366 | 215 |
| | Accidents increasing | 47 | 23 |
| | Total | 2422 | 1376 |

The basis of estimating these benefits is the estimation of the values of each cost factor and extent of reduction as the benefits of the plan. Thus, implementation of the three options will be discussed, including the "status quo", "directional interchange" and " standard cloverleaf interchange". These options were implemented in the Tehran transportation system simulation model with EMME software in the years after the operation. Based on the results presented in the above tables and comparing the two options of the standard cloverleaf and directional interchanges, the advantages of using these interchanges in the network in each of the studied years are provided in **Fig. 4**, **Fig. 5**, **Fig. 6**, and **Fig. 7**.

In **Fig. 4**, **Fig. 5**, **Fig. 6**, and **Fig. 7**, the comparison between benefits of using directional and cloverleaf interchanges is illustrated in different operation years for each of the benefits: reducing travel time, reducing fuel consumption, reducing accidents, and reducing pollutants.

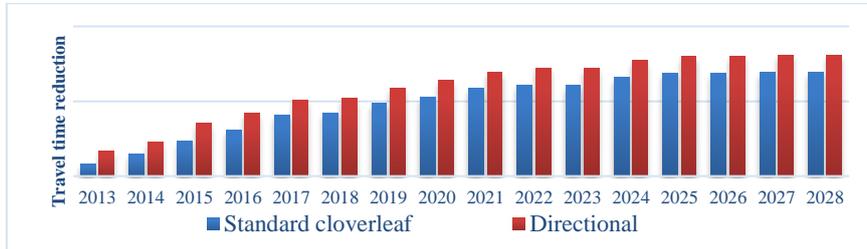

**Fig. 4**. Travel time reduction in different years

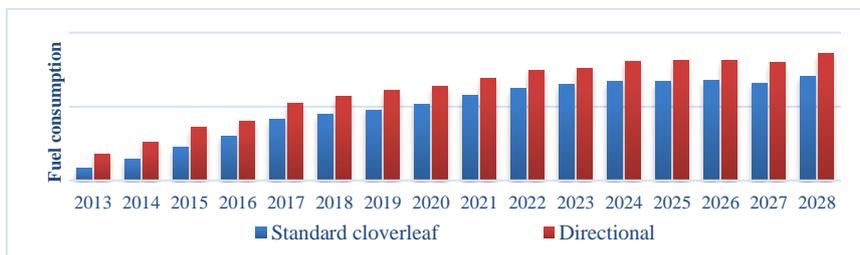

**Fig. 5.** Fuel consumption reduction in different years

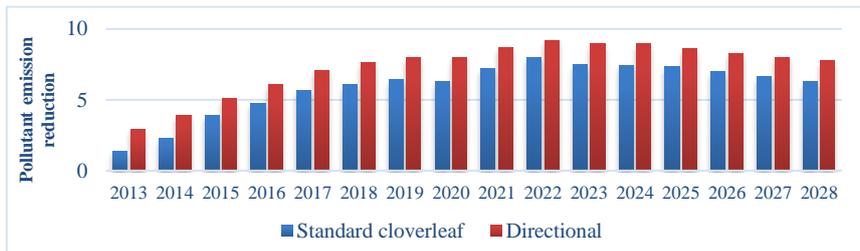

**Fig. 6.** Pollutant emission reduction in different years

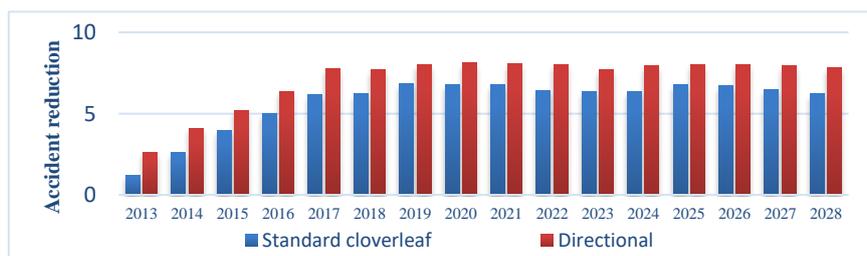

**Fig. 7.** Accident reduction in different years

In the comparison between the two proposed options for geometric promotion of the interchange, the directional interchange is 28% better than standard cloverleaf interchange. Also, during the construction of interchanges, this superiority is maintained. This shows that despite the initial higher cost, the directional option has higher interests during the operation. Comparison of costs and interests of directional and cloverleaf interchange in different years can be seen in **Fig. 8** and **Fig. 9**, based on the Net Present Value index. According to **Fig. 8** and **Fig. 9**, both plans have a non-negative "net present value" in the third year. Since the profitability of the project is before the end of the plan's useful life, therefore, the construction and exploitation of the studied interchanges are economically justified. It can be proven that geometric promotion in highway interchanges has a strong justification from an economic standpoint, safety, and traffic.

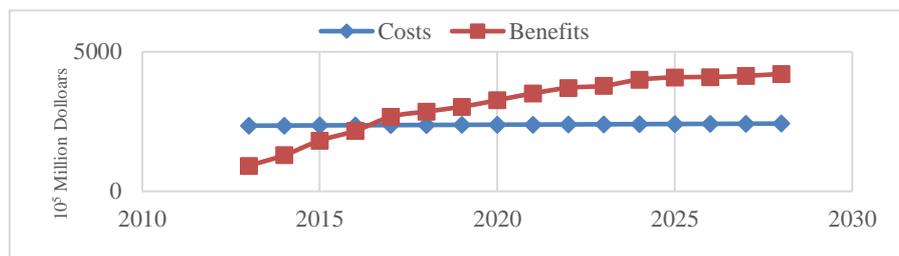

**Fig. 8.** Costs and benefits of directional interchange plan

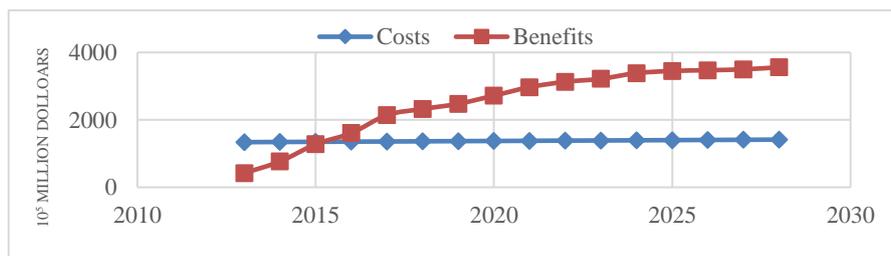

**Fig. 9.** Costs and benefits of the cloverleaf interchange plan

## 4  Conclusion

There have been two scenarios for improving the geometry characteristics of the interchange. The first one has been for increasing the capacity and transforming the status quo to improve cloverleaf interchange. The second one has been directional exchange with the three-story bridge design. **Table 16** displays the process and analysis results of EMME software.

There are two methods for geometric improving the interchanges including line capacity increasing and changing the interchange to another interchange with a higher capacity. With geometric improving the cloverleaf interchange, it is concluded that method 2 has 27**%** economic superiority. The directional interchange plan has a $B/C = 1$ between the second and third years. After that, with much higher benefits than costs, it will have a very favorable impact on the network. The directional interchange plan has a $B/C = 1$ between the third and fourth years. Benefits increase with a lower slope than the directional option. A sensitivity analysis indicated that the fluctuations in the fuel price are most effective in improving geometric interchanges (coefficient: %55). Also, travel costs, accidents costs, and environmental costs are 38%, 4%, and 3% respectively.

Table 16. The process and analysis results of EMME/2 software

| Type | Reduction of travelled distance (%) | Reduction of fuel consumption (million litter) | Reduction of air pollution (%) | Reduction of travel time (million hours) |
|---|---|---|---|---|
| Directional | 2.67 | 74 | 2.88 | 16 |
| Cloverleaf | 1.25 | 33 | 1.25 | 7 |